\documentclass{nature}

\usepackage{graphicx,amsmath}
\usepackage{booktabs}
\usepackage[flushleft]{threeparttable}

\title{3D spherical-cap fitting procedure for (truncated) sessile nano- and micro-droplets \& -bubbles}

\author{Huanshu Tan$^{1}$, Shuhua Peng$^{2}$, Chao Sun$^{3,1}$, Xuehua Zhang$^{2,1}$, Detlef Lohse$^{1,4}$}

\begin{document}

\maketitle

\begin{affiliations}
 \item Physics of Fluids group, Department of Science and Technology, Mesa+ Institute, and 
J. M. Burgers Centre for Fluid Dynamics, University of Twente, P.O. Box 217, 7500 AE Enschede, The Netherlands,
 \item Soft Matter \& Interfaces Group, School of Engineering, RMIT University, Melbourne, VIC 3001, Australia,
 \item Center for Combustion Energy \&
Department of Thermal Engineering, Tsinghua University, China,
 \item Max Planck Institute for Dynamics and Self-Organization, 37077 G\"ottingen, Germany.
\end{affiliations}

\begin{abstract}
In the study of nanobubbles, nanodroplets or nanolenses immobilised on a substrate, a cross-section of a spherical-cap is widely applied to extract geometrical information from atomic force microscopy (AFM) topographic images.
In this paper, we have developed a comprehensive 3D spherical cap fitting procedure (3D-SCFP) to extract morphologic characteristics of complete or truncated spherical caps from AFM images.
Our procedure integrates several advanced digital image analysis techniques to construct a 3D spherical cap model, from which the geometrical parameters of the nanostructures are extracted automatically by a simple algorithm. The procedure takes into account all valid data points in the construction of the 3D spherical cap model to achieve high fidelity in morphology analysis.  We compare our 3D fitting procedure with the commonly used 2D cross-sectional profile fitting method to determine the contact angle of a complete spherical cap and a truncated spherical cap. The results from 3D-SCFP are consistent and accurate, while 2D fitting is unavoidably arbitrary in selection of the
 cross-section and has a much lower number of data points on which  the fitting can be  based, which in addition
 is biased to the top of the spherical cap. 
 We expect 
 that the developed 3D spherical-cap fitting procedure will find many applications in imaging analysis. 
\end{abstract}

\section{Introduction}
\label{intro}
Atomic force microscopic (AFM) imaging has been one of the most popular techniques in the study of surface nanobubbles, nanodroplets and many other spherical cap nanostructures \cite{lohse2015rmp,lou2000,ishida2000,zhang2007nanodroplet}.
Thanks to the high spatial resolution of AFM measurements, the size of nanobubbles and nanodroplets immobilised on the substrate can be characterised accurately in all three dimensions. Geometrical features of the spherical cap structures, such as lateral extension of the footprint, height and contact angle, can be compared on a quantitative level \cite{lohse2015rmp,lohse2015,german2014,wang2002evaluation}.  Although most of those parameters are directly read out from the images, the contact angle of nanobubbles and nanodroplets requires further imaging analysis and data processing. As surface nanobubbles or nanodroplets are usually spherical caps due to dominance of capillarity on such small dimension, the data analysis requires the construction of an ideal spherical cap to fit the AFM data. 

Up to now, fitting a two-dimensional (2D) cross-sectional profile  is the most common approach for determining the contact angle of nanobubbles and nanodroplets \cite{wang2002evaluation,zhang2006langmuir,zhang2008langmuir,simonsen2004,uddin2011novel,wang2015,zhang2012morphology,xu2014,peng2015spontaneous}.
With an AFM off-line software, the cross sectional profile through the centre of bubbles or droplets can be conveniently read out from the image. For instance, Simonsen et al. \cite{simonsen2004} and Wang et al. \cite{wang2015} calculated the contact angle of nanobubble, based on the measured height and footprint size in the cross-section. Wang et al. \cite{wang2002evaluation} determined the droplet size by simply averaging the values calculated from several different cross-section profiles. In a slightly improved method, the cross-sectional profile is fitted with a portion of an ideal circle (i.e. an arc). The angle subtended by the arc and the flat baseline is the contact angle of the measured nanostructures. This method has been applied by Zhang et al. \cite{zhang2006langmuir,zhang2008langmuir,xu2014,peng2015spontaneous} and Yang et al. \cite{yang2003} to obtain the contact angle of nanobubbles and nanodroplets on several flat substrates.  However, the drawbacks from 2D fitting are the arbitrary selection of the cross-section, which leads to large variation in the contact angle measured along different directions in the image \cite{wang2002evaluation,uddin2011novel,mugele2002,antonio2009}.

To obtain a contact angle accurately, the contribution of \textit{all valid} the data points on the nanostructure must be taken into account in the reconstruction of the spherical cap model. This requires the development of a comprehensive 3D fitting \cite{borkent2010,song2011}. 
Although Song et al. \cite{song2011} applied a 3D spherical-cap fitting to analyse the morphology of nanobubbles, sensible exclusion of  unreliable data points is an important aspect in 3D fitting which has not been considered so far.
Particularly for highly curved bubbles or droplets,  the data points close to the three-phase contact line should be excluded  (i.e. contact angle larger than $90^o$), due to the tip-sample convolution \cite{lohse2015rmp,garcia2002}.  In the analysis of an image of very small bubbles and droplets, there may be also potential effects from the disjoining pressure \cite{chengara2004,sibley2014}, and hence only the data points above a certain threshold are valid for the fitting \cite{borkent2010,song2011}.  
Furthermore, no 3D fitting procedure has been reported for the analysis of a \textit{truncated} spherical cap on a substrate with physical structures.
A simple example of truncated droplets is that the droplets form at the rim of a microcap \cite{peng2015spontaneous}. The shape of those truncated droplet resembles an apple slice with a part bitten off from the flat face.  

In this paper, we have developed a 3D spherical-cap fitting procedure (3D-SCFP). 
The procedure employs the techniques of digital image processing to recognise the features from a spherical cap.  It performs well for an isolated nanodroplet sitting on a flat substrate, but also for a truncated nanodroplet on the rim of a microcap \cite{peng2015spontaneous,peng2016}.
In the latter case, a feature extraction method, the Circle Hough Transform, is performed to distinguish the AFM data points from the part of the underlying microcaps and from the truncated nanodroplets \cite{mathai2015wake,prakash2012gravity,ballard1981generalizing}.

\begin{figure*}
\centering
\resizebox{\linewidth}{!}{%
  \includegraphics{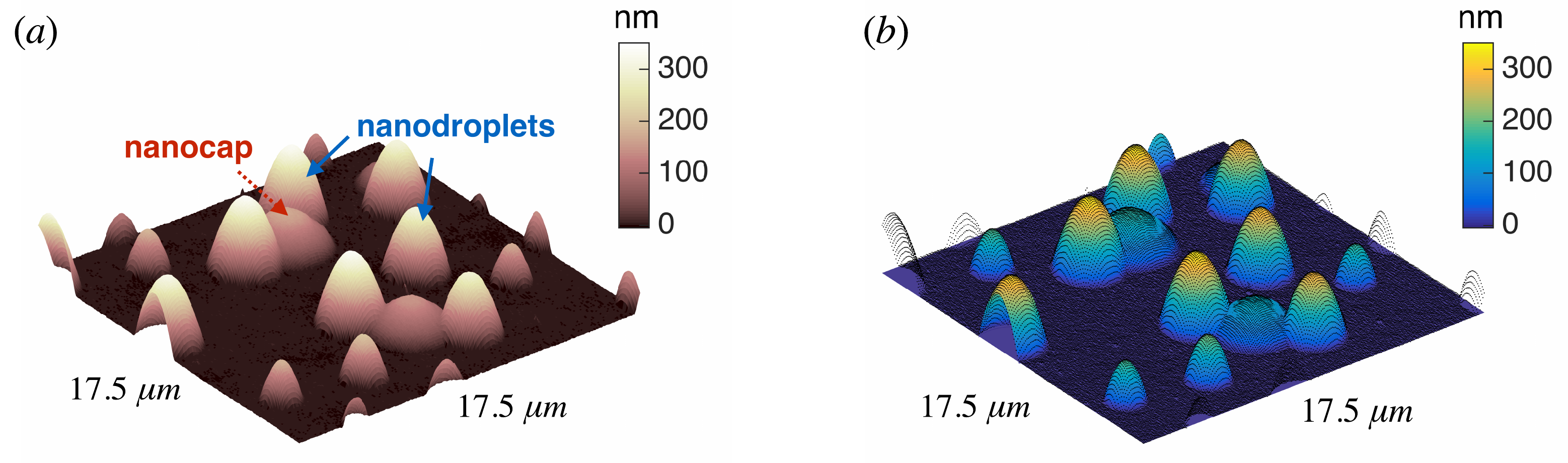}
}
\caption{Three-dimensional representation of the nanodroplets and overlapped nanocaps (the arrows). (a) AFM image of nanodroplets sitting on the flat substrate and on the rim of nanocaps. Image size: $17.5 \mu m \times 17.5 \mu m$. (b) Reconstructed three-dimensional image from the 3D spherical-cap fitting procedure. The AFM data points are displayed as the black dots, showing a nearly perfect agreement with the fitting surface.}
\label{fig:AFM}       
\end{figure*}

With our proposed 3D-SCFP, we will analyse polymerised nanodroplets in AFM images collected from our previous work ~\cite{peng2016}, as represented in figure~\ref{fig:AFM}a. The nanodroplets (indicated by blue solid arrows) to be analysed include both complete spherical caps on a flat area and truncated droplets by the underlying microstructures (indicated by red dashed arrow).
3D-SCFP will translate AFM data points to ideal spherical caps that nicely match the nanodroplet morphology, as displayed in figure~\ref{fig:AFM}b.  For those truncated nanodroplets, the procedure recognises the data points from a part of the spherical cap and reconstruct a whole spherical cap model.  

The following sections of the paper are organised as follows: Section 2 details the algorithm of 3D-SCFP, and section 3 discusses the influence of an important parameter, the threshold of height cut-off, on the 3D fitted result. 
After that, the paper gives two examples in section 4, showing the comparison between 3D-SCFP and 2D cross-sectional profile fitting method.
The results reveal the 3D-SCFP is robust, compared to the 2D fit method.  The Matlab codes of 3D-SCFP are provided in the supplementary material, with free access to the readership.

\section{Detailed 3D fitting procedure}
\label{sec:1}
The algorithm consists of four parts: raw image preprocessing, objective detection, objective identification, and cut-off of vertical data points near the rim.

\begin{figure*}
\centering
\resizebox{0.9\linewidth}{!}{%
  \includegraphics{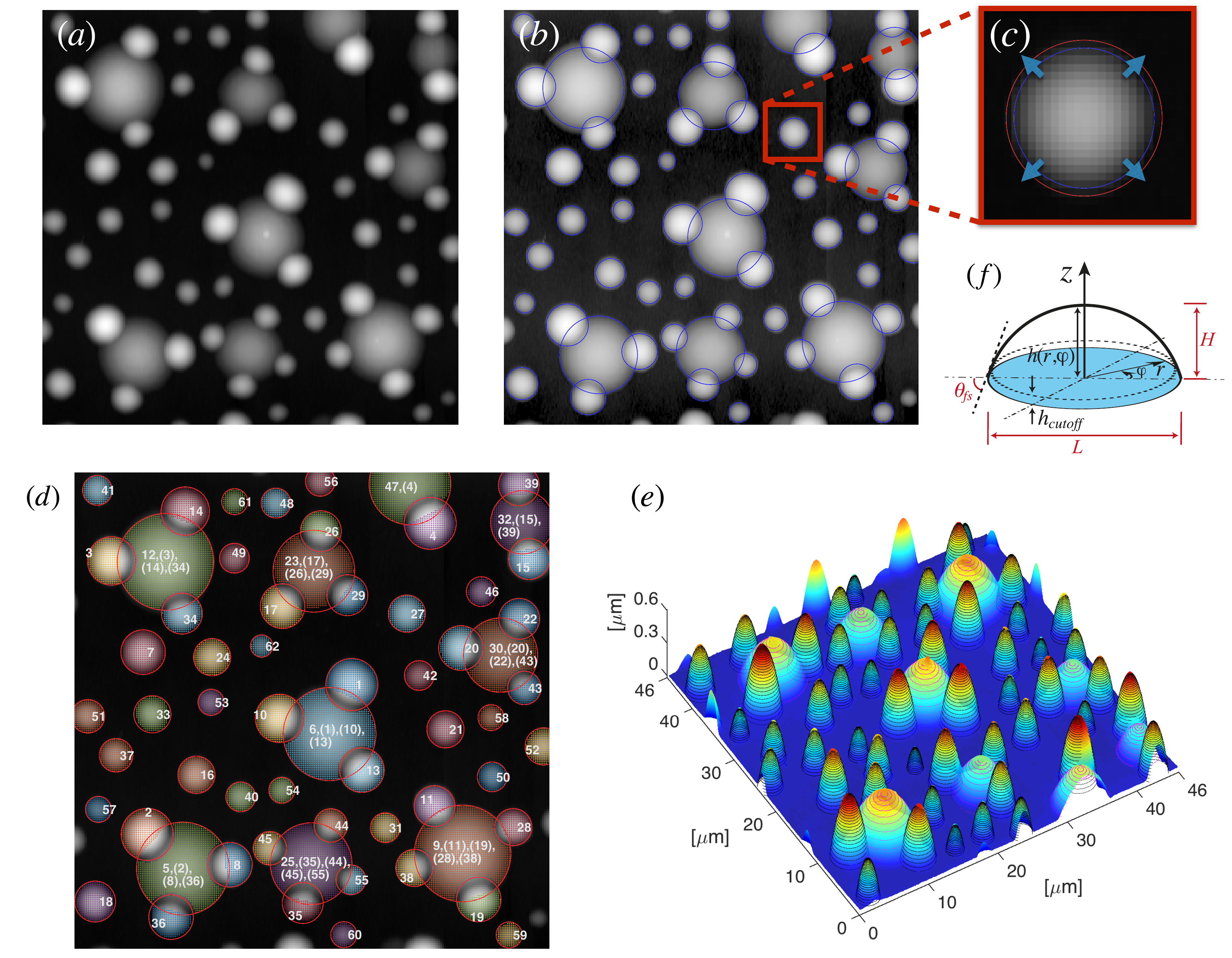}
}
\caption{Screen snapshot series of the 3D spherical-cap fitting procedure (a-e). (a) 2D representation of an AFM sample image. The pixel intensities of the graphics correspond to the sample height. (b) In the second step of the procedure, the footprints of the objectives are detected as blue circles. In order to surround all the data points (pixels) of each objective, the radii of the circles are designed to be adjustable as shown in (c). 
(d) shows the identification of the objective with coloured data points for 3D fit in the final step. The relation between all the nanodroplets and the overlapped microcaps is also recognised. (f) The final fitted ideal geometries (the dotted lines), showing a very good agreement with corresponding AFM sample image displayed in 3D (the coloured surface).
(d) The sketch of an isolated nanodroplet with annotations. The $h_{cutoff}$ is defined as the height above the flat substrate.}
\label{fig:details}       
\end{figure*}
\subsection{Raw image pre-processing}
The AFM images pretreated by second-order flattening by AFM off-line software are used in this step. A representative image is shown in $I$ (Fig.~\ref{fig:details}a). 
We create a so-called gamma encoded image $I'$ (Fig.~\ref{fig:details}b) from the raw image $I$. 
Gamma encoding is defined as $V_{out}=AV_{in}^{\gamma}$ with $1>\gamma>0$ \cite{reinhard2010high}, where $V_{out}$ and $V_{in}$ present pixel intensity of images $I'$ and $I$, respectively, and $A$ is a constant.
The operation was implemented through ``\textit{imadjust}'', a basic function in MATLAB. 
After the operation, the pixel intensity of the whole image $I'$ is nonlinearised, resulting in a sharper contact line with the flat substrate. The obtained image is shown in figure~\ref{fig:details}b. The sharp boundary in the image increases the reliability of the feature detection in the next step \cite{najafabadi2011novel}.

\subsection{Objective detection}

This step determines the circular footprint of a complete nanodroplet on the flat substrate, but also the footprints of multiple truncated nanodroplets and the underlying microcaps (blue circles in Fig.\ref{fig:details}b). 
An edge detection technique, namely the Circle Hough Transform \cite{mathai2015wake,prakash2012gravity,ballard1981generalizing}, is applied to the gamma encoded image $I'$.
The Circle Hough Transform is a feature extraction technique for circle detection with robustness in the presence of noise and occlusions \cite{davies2012computer}.
In order to have all the data points of a complete droplet included in the circle of the footprint, we set a parameter $pa_0$ to adjust the radii of the fitting. 
As demonstrated in the zoom-in in figure~\ref{fig:details}c, the initially detected circle (blue) does not enclose all the data points of a nanodroplet. By increasing the radius, a new circle (red) is generated, now wrapping up all data points (pixels) in the image. The rim of the truncated nanodroplet on the flat substrate is part of a circle, which together with the rim of the underlying nanocap can be detected through the advanced feature extraction method, i.e., the Circle Hough Transform.  As shown in figure~\ref{fig:details}b, those blue connected rings fit the boundaries of the nanocap and the truncated droplets nicely. 

\subsection{Objective identification} 
If there are only isolated nanodroplets in the AFM image, the step discussed in this subsection is not essential.
For the case with nanodroplets sitting on the rim of a microcap, it is crucial to automatically identify whether the detected objective is a nanodroplet or a microcap. 
The identification is based on the characteristics from experiments: (i) Nanodroplets can sit above microcaps, while the reverse does not apply; (ii) Serval nanodroplets can nucleate on the rim of an identical microcap;
(iii) The truncated nanodroplets are higher or lower than the microcaps, depending on their materials (details refer to Ref.~\cite{peng2016}). 

Based on those information, we perform some conditional statements to determine the nanodroplet and the microcap.
Then the data points in the overlap region are attributed to the recognised nanodroplets according to characteristics (i) (see appendix A for the details).
In figure~\ref{fig:details}d, the identification numbers are displayed. 
Behind the label number of each microcap, the label numbers of its overlapping nanodroplets are also listed in the parentheses.

\subsection{Cut-off of vertical data points near the rim}
As already mentioned in the Introduction, the data points near the rim are subject to the influence of tip convolution and the disjoining pressure. 
Hence, an appropriate cut-off of the droplet rim morphology is needed for proper profile fitting, no matter 2D or 3D fitting method is applied.
Here we take those data points above a threshold height for 3D spherical cap fitting (details are discussed in section \ref{sec:2}). 
The threshold is a new parameter, $h_{cutoff}$ (sketch in Fig.\ref{fig:details}f). The valid data points are shaded in different colours in figure~\ref{fig:details}d, and fitted with a spherical cap by the program. The fitting is generated by minizing the cost function, which is the summed distance from the data points to the ideal spherical surface. 
The system of the nonlinear equation for the optimisation is directly solved by MATLAB.
As shown in figure~\ref{fig:details}e, the whole topographic AFM image (the coloured surface) is perfectly matched by the corresponding ideal spheres (the black dot lines for nanodroplets and magenta for microcap).
The fitting results are saved as ideal spheres' radii and their centre-point coordinates with respect to the flat substrate surface.
Based on these parameters, the calculations of footprint lateral extension $L$, height $H$, contact angle on flat substrate $\theta_{fs}$ (Fig.\ref{fig:details}f), contact angle on the nanocap $\theta_{mc}$ and other relevant geometrical parameters are determined mathematically.

\begin{figure*}
\centering
\resizebox{0.95\linewidth}{!}{%
  \includegraphics{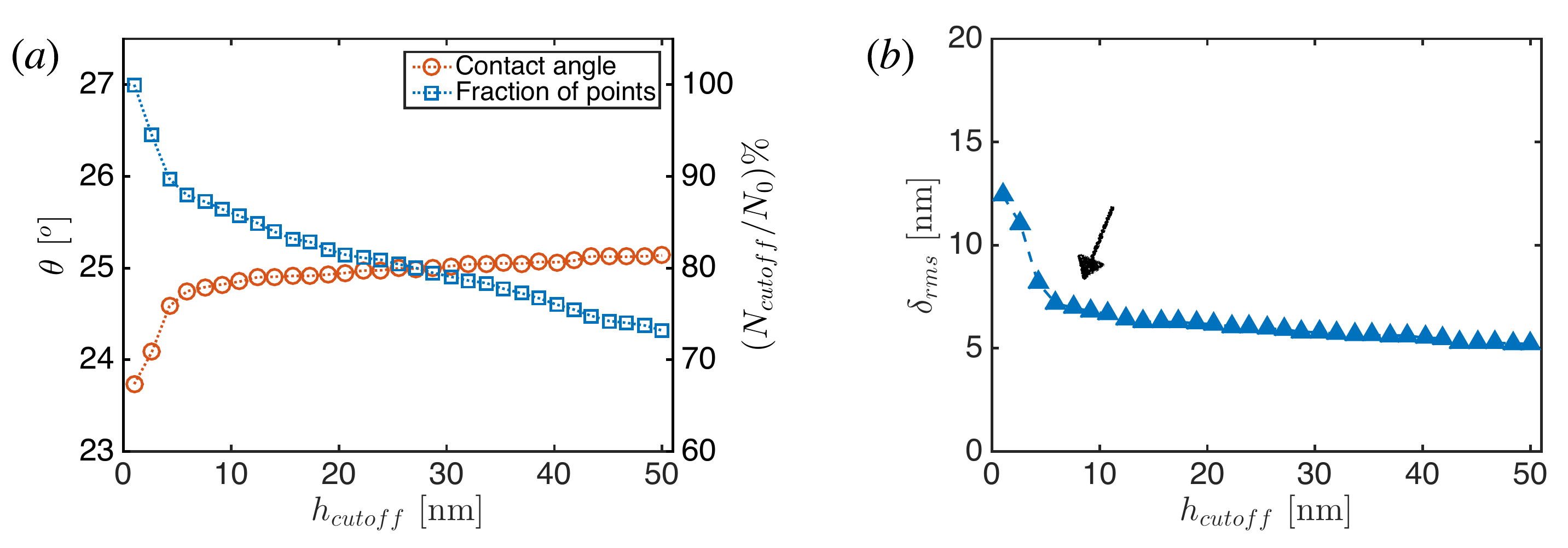}
}
\caption{(a) The contact angle and the fraction of the corresponding fitted data points versus the cut-off height $h_{cutoff}$ in 3D spherical-cap fit. 
When $h_{cutoff}$ is smaller than a certain value (around 8 nm marked by an arrow), the fitted contact angle has a relatively large variation. 
(b) The fit error $\delta_{rms}$ versus the cut-off height $h_{cutoff}$ in 3D spherical-cap fit.
The fit error $\delta_{rms}$ has a sudden increase when $h_{cutoff}<8$ nm,
indicating that the data points below 8 nm are questionable for 3D spherical-cap fit.
The 3D representation of the nanodroplet is shown in figure~\ref{fig:res1}(a).
}
\label{fig:hcutoff}       
\end{figure*}

\section{Evaluation of the effects from cut-off threshold}
\label{sec:2}

\begin{figure*}
\centering
\resizebox{0.95\linewidth}{!}{%
  \includegraphics{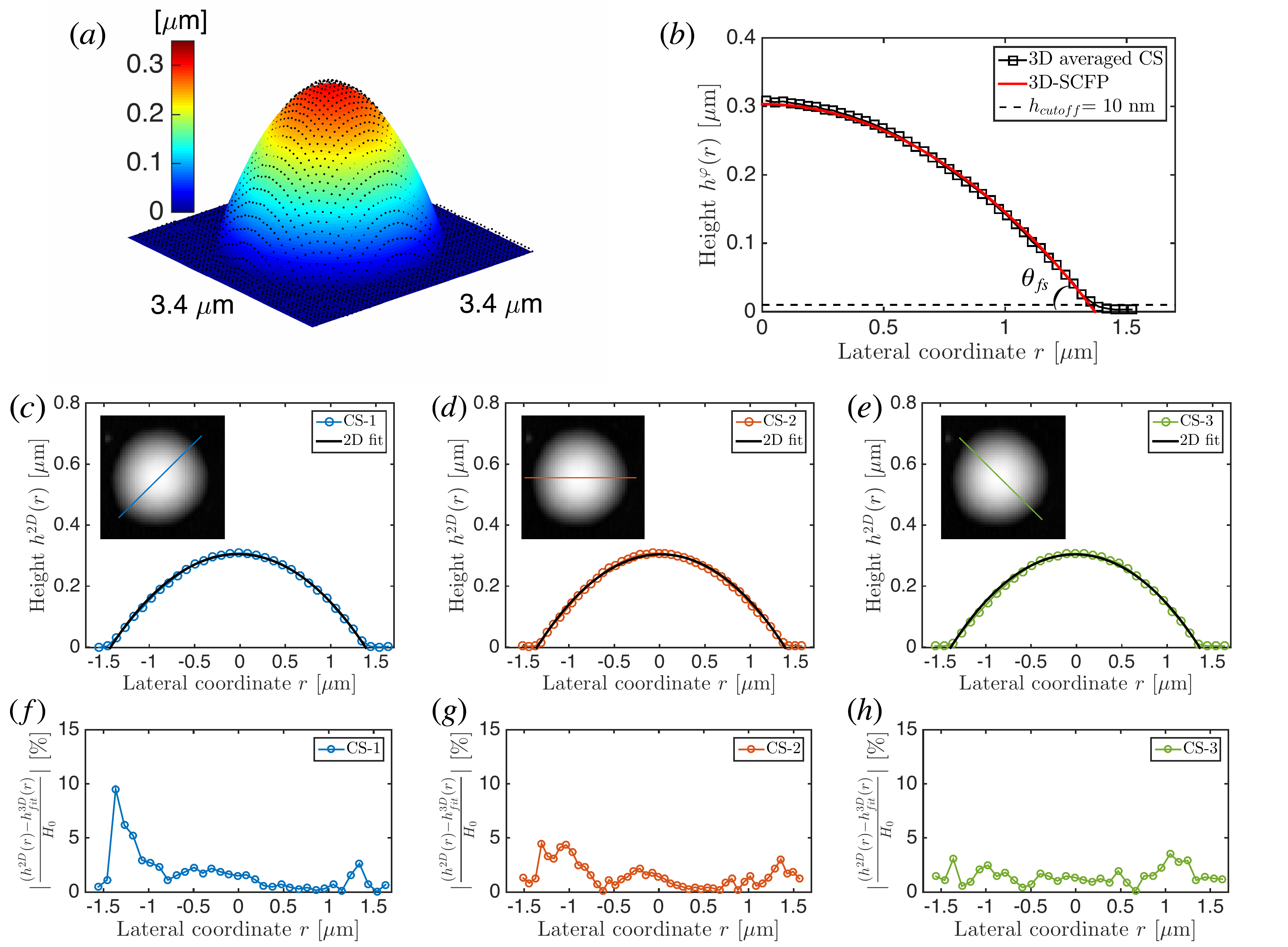}
}
\caption{(a) Reconstructed 3D image of an AFM sample image with an isolated nanodroplet sitting on the flat substrate. The AFM data points displayed as the black dots fit well with the coloured ideal spherical surface calculated from 3D-SCFP.  Image size: 3.4 $\mu m$ $\times$ 3.4 $\mu m$. Droplet height: 0.3 $\mu m$.
(b) Map of residuals for 3D fitting with definition as $|h(r,\varphi)-h_{fit}^{3D}(r,\varphi)|/H_0$. The residuals are coded by colour.
(c) Defined averaged cross-sectional profile $h^{\varphi}(r)$ (square), showing an excellent agreement to the 3D fitting result (red curve).
(d-f) Three cross-sectional profiles CS-1, CS-2 and CS-3 of the same nanodroplet extracted along different directions. The cross section cutting lines are shown in the inserted AFM images, respectively. Based on the spherical-cap assumption, the fitting shapes (the black curves) match the three different cross-sectional profile (circle) very well, but give different fitted contact angles (Tab.~\ref{tab:res}).
Corresponding to (d-f), (g-i) show the deviation of each profile $h^{CS}(r)$ from its ideal spherical-cap fitting result $h_{fit}^{3D}(r)$, which considers the contributions of all the data points.}
\label{fig:res1}       
\end{figure*}

 In this section, we show the effect of the threshold by analysing the droplet in Fig.~\ref{fig:res1}a.
The nanodroplet is analysed by 3D-SCFP with 31 different $h_{cutoff}$ from 1 nm up to 30 nm (around 10\% height of the nanodroplet), while all the other fitting parameters are fixed.
With an increase in the threshold $h_{cutoff}$, the fraction of data points above the threshold decreases. 
The obtained contact angles from 3D-SCFP with different given $h_{cutoff}$ are plotted in figure~\ref{fig:hcutoff}a, showing a strong dependence on $h_{cutoff}$ for $h_{cutoff}<8$ nm. 
We define the fit error $\delta_{rms}$ as root mean square of the radial distance between the data points and the ideal sphere.
The plot of the error versus the threshold in figure~\ref{fig:hcutoff}b shows that inclusion of the data points close to the substrate leads to derivation of the fitting from an ideal sphere.
Too small $h_{cutoff}$ causes large 3D spherical-cap fitting error, while too larger $h_{cutoff}$ leads to less contributions from valid data points. 
An appropriate cut-off is determined to be the point just after the turning point in the plot of figure~\ref{fig:hcutoff}b, marked by an arrow.
For each group of AFM images from a certain experimental condition, the ``turning points'' was determined by looping the procedure.

\section{3D-SCFP versus 2D fitting}
\label{sec:4}
We provide in this section the comparison between the results of 3D-SCFP and 2D cross-sectional profile fitting method in two cases: an isolated nanodroplet and two truncated nanodroplets. 

\subsection{Case 1: An isolated nanodroplet on the flat substrate}

The threshold $h_{cutoff}$ for the fitting is set to be 8 nm, which is determined in the way described in section 3 (Fig.~\ref{fig:hcutoff}a).
In figure~\ref{fig:res1}a, the 3D fitting result (the coloured spherical surface) shows an excellent agreement with the AFM data points (the black dots). 
To quantify the error of 3D fitting, the residual is defined as:
\begin{equation}
|h(r,\varphi)-h_{fit}^{3D}(r,\varphi)|/H_0.
\label{eq:re}
\end{equation}
 The height of each point
$h(r,\varphi)$ is expressed as function of  the 
 cylindrical coordinates (sketch Fig.~\ref{fig:details}f). $h_{fit}^{3D}(r,\varphi)$ is the corresponding ideal 3D profile by 3D fitting. The absolute difference is normalised by the droplet height $H_0$.
As displayed in the residual map (Fig.~\ref{fig:res1}b), the residual within most region is less than 1\%, which is comparable to the signal noise.
The maximum deviation, around 8\%, occurs at the region near the contact line.
From 3D fitting result, the contact angle $\theta_{fs}$ is calculated and listed in Table~\ref{tab:res}.

We introduce an averaged cross-sectional profile $h^{\varphi}(r)$ defined as:
\begin{equation}
h^{\varphi}(r)=\frac{1}{2\pi}\int_{0}^{2\pi}h(r,\varphi)d\varphi,
\label{eq:ap}
\end{equation}
where $h(r,\varphi)$ is the height of the data point in the cylindrical coordinates (sketch Fig.~\ref{fig:details}f).
The origin of the coordinate system is the centre point of the circular footprint of the ideal sphere on the flat substate.
As displayed in figure~\ref{fig:res1}c, the 3D spherical-cap fitting result (the red line) and the averaged cross-sectional (the black square line) profile perfectly match with each other as expected.

In 2D fitting,  the nanodroplet is assumed to be an ideal spherical cap. Part of three circles are used to fit three different cross-sectional profiles of the same nanodroplet: CS-1 (along $45^o$ direction), CS-2 ($0^o$) and CS-3 ($135^o$), as shown in figures~\ref{fig:res1}d, e and f, respectively.
The data for the fitting are selected as the points that are higher than 8 nm, which are displayed as solid dots in figures~\ref{fig:res1}d-f.
The fitting results show that the 2D fitting circle (the black line) and the profile data (the circular dots) in each cross section match well.
The contact angles fitted from those three profiles data are listed in Table~\ref{tab:res}. 
$\theta_{fs}$ obtained from CS-1 is smaller than the fitted results of other two cross-sectional profiles. 
We display the residuals along different cross-section directions, i.e. $|h^{CS}(r)-h_{fit}^{3D}(r)|/H_0$, in figure~\ref{fig:res1}g-i.
The deviation profiles vary along different cross-section directions and can be asymmetry.
Hence, different selections of the direction of the cross-section in 2D cross-sectional profile fitting give inconsistent results, as also concluded 
from the fitted contact angles shown in Table~\ref{tab:res}.
The inconsistency of the 2D fitting results has already been noticed by other researchers \cite{uddin2011novel,mugele2002,antonio2009}.
Such difference among different cross-sections might be associated with different extent of experimental errors along different directions of the image: longer scanning time required to collect the data points across different scan lines and along the same line and hence the cross-section along the scan line in (Fig.~\ref{fig:res1}g) is more accurate.
Another possibility is that the droplet is never an ideal spherical cap, due to the unavoidable chemical heterogeneities on the surface \cite{bonn2009,stauber2014}. 
If so, the heterogeneity may lead to the irregularity on the base boundary and a non-circular droplet footprint.
Overall, the variation of the extracted cross-sectional profile jeopardise the reliability of 2D cross-sectional profile fitting method.

\begin{table*}
\centering
\caption{Comparison of the calculated contact angles through 3D-SCFP and 2D cross-sectional profile fitting method. The cross section labels correspond to those in figures~\ref{fig:res1} and \ref{fig:res2}. Two kinds of contact angle are compared: $\theta_{fs}$, contact angle of nanodroplets or nanoocap on flat substrate, and $\theta_{mc}$, contact angle of the nanodroplet on a nanocap.
}
\label{tab:1} 
\begin{tabular}{l  c c c c c c  c}
\hline\noalign{\smallskip}
  & CS-1 & CS-2  & CS-3 & CS-4  & CS-5  & CS-6a,b,c &3D fit\\
\noalign{\smallskip}\hline\noalign{\smallskip}
\quad $\theta_{fs}$            & 24.63$^o$& 25.38$^o$   & 25.34$^o$  & $ $             & $ $                &   $ $                    & 24.82$^o$\\
a: $\theta_{fs}$ (drop)       & $ $             & $ $             & $ $            & 29.08$^o$   & $ $               &   27.32$^o$         & 26.99$^o$\\
a: $\theta_{mc}$ (drop)     & $ $             & $ $             & $ $            & 30.73$^o$   & $ $               &   $ $                     & 28.87$^o$\\
b: $\theta_{fs}$ (drop)       & $ $             & $ $             & $ $            & $ $              &  27.61$^o$   &      28.16$^o$       & 27.82$^o$\\
b: $\theta_{mc}$ (drop)     & $ $             & $ $             & $ $            & $ $              &  29.42$^o$   &       $ $                 & 30.28$^o$\\
c: $\theta_{fs}$ (cap)         & $ $             & $ $             & $ $            & 7.16$^o$    & 7.45$^o$      &    7.81$^o$          & 7.65$^o$\\
\noalign{\smallskip}\hline
\end{tabular}
\label{tab:res}
\end{table*}

\subsection{Case 2: Truncated nanodroplets sitting on the rim of a nanocap}
\begin{figure*}
\centering
\resizebox{0.95\linewidth}{!}{%
  \includegraphics{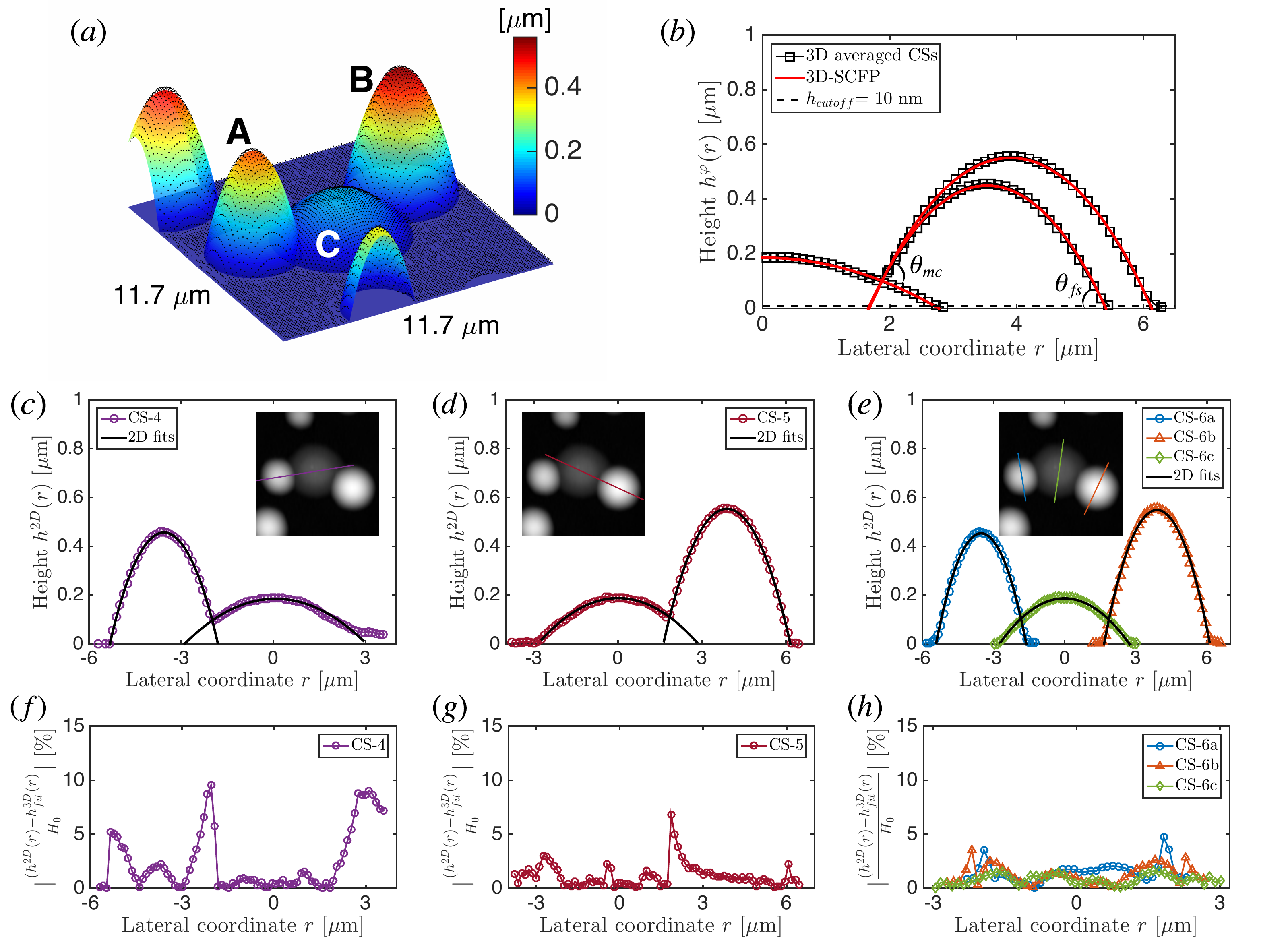}
}
\caption{(a) Reconstructed 3D image of an AFM sample image, in which two nanodroplets A and B sitting on the rim of a nanocap C. The AFM data points displayed as the black dots fit well with the coloured ideal spherical surface calculated from 3D-SCFP. 
(b) Map of residuals for 3D fitting.
(c) Defined averaged cross-sectional profiles $h^{\varphi}(r)$ (square), showing an excellent agreement to the 3D fitting results (red curves).
The positions of the nanodroplets and the nanocap are based on the centre distances.
(d-i) hold the same definitions as in figure~\ref{fig:res1}(d-i) (the calculated contact angles refer to Tab.~\ref{tab:res}).
}
\label{fig:res2}       
\end{figure*}

We now analyse the image with two truncated nanodroplets A and B sitting on the rim of a nanocap C (Fig.~\ref{fig:res2}a).
The threshold $h_{cutoff}$ for nanocap and nanodroplets is set to be 10 nm and 20 nm, respectively.
3D-SCFP automatically distinguishes the data points on nanodroplets A, B and on the nanocap C, respectively, and then fits valid data points of each them to an ideal sphere.
As shown in figure~\ref{fig:res2}a, the fitting result is displayed in coloured spherical surface and compared with AFM data points in black dots.
A residual map of 3D fitting defined as in equation~(\ref{eq:re}) is given in figure~\ref{fig:res2}b ($H_0$ for the map is the height of nanodroplet B here).
The high performance of 3D fitting is thus quantitively verified.
The maximum residual is less than 10\% and is located at the contact region between the nanodroplets and the nanocap. 
The map here also exposes the inconsistency between different cross-sections.
Based on the definition in equation~\ref{eq:ap}, we also build their averaged cross-sectional profiles.
Three averaged cross-sectional profiles (black square) are integrated in a height-radial coordinate with the origin defined as the footprint centre of the nanocap C  (Fig.~\ref{fig:res2}c).
The 3D fitting results are displayed as the red lines, showing perfect agreement to the averaged profiles.
Here, we introduce the contact angle of the nanodroplet on the overlapped nanocap, labeled as $\theta_{mc}$.
Then both $\theta_{fs}$ and $\theta_{mc}$ are calculated from the 3D fitting results and are listed in Table~\ref{tab:res}.

For this case, it is no longer convenient to apply 2D cross-sectional profile analysis any more.
To obtain the correct contact angle $\theta_{mc}$, we must pick up the cross section that passes both the centre point of the nanodroplet and that of the nanoocap. 
However, there are unavoidable artificial errors of locating the centra positions of the nanodroplet and the nanocap.
When the cutting line of the selected cross section is not in the column direction or row direction of the image pixel matrix, the profile of an selected cross section can not be precisely depicted, because the cutting line may not pass pixel centres (as sketched in Fig.~\ref{fig:appB} in appendix B).
In low pixel resolution image, this problem is even more crucial.
In order to reduce this influence as much as possible in this section, firstly we choose an AFM image with a relatively high resolution.
The lateral dimension of a nanodroplet is depicted by more than 40 pixels (Fig.~\ref{fig:res2}a).
Secondly, the extraction of the cross-sectional profile in a particular direction is improved by applying a linear fitting, as described in appendix B.

In figures~\ref{fig:res2}d, e and f, two cross sections (CS-4 and CS-5) that pass two centre points of the objectives and three complete cross-sectional profiles of each objective (CS-6a, b and c) are displayed.
Notably, the profile of nanocap C in CS-4 is not only overlapped by nanodroplet A, but also slightly by the edge of nanodroplet B, which is not conspicuous.
The overlap reduces the number of the valid data points for a 2D fit.
This situation tends to be common when there are more nanodroplets sitting on a same nanocap. 
We only apply 2D fits to the valid segment of the extracted profiles that are 
higher than 10 nm for nanocap C and higher than 20 nm for nanodroplets A and B.
As shown in figures~\ref{fig:res2}d-f, the 2D fitting results (the black circles) have a good agreement to the corresponding profile data (solid dots).
Based on the 2D fitting results, $\theta_{fs}$ and $\theta_{mc}$ of the nanodroplets and the $\theta_{fs}$ of the nanocap are calculated and are listed in Table~\ref{tab:res}.
Again, the normalised deviation of $h^{CS}(r)$ from $h_{fit}^{3D}(r)$ is calculated and is displayed in figures~\ref{fig:res2}g-i.
It shows that the deviation increases in the profile part nearby the contact point/line.
Especially for nanodroplet A in CS-4, its profile severely departs from its 3D fitting profile (Fig.~\ref{fig:res2}g),
leading to 2$^o$ difference for both $\theta_{fs}$ and $\theta_{mc}$ (Tab.~\ref{tab:res}).
However, with the partial profile of nanodroplet B in CS-5, the difference of fitted results of the 2D method to that of the 3D-SCFP is less than 1$^o$.
We also acquired similar fitting result $\theta_{fs}$ when applying 2D fits to the complete extracted profiles CS-6a, b and c.
But in general the contact angle $\theta_{mc}$ is unaccessible from the
complete profiles as in general the cross sections do not go through the centres of both the nanodroplets and the nanocap.
Thus again the  3D-SCFP fit demonstrates its superiority in this complex case.

\section{Conclusion and Outlook}
\label{sec:5}
In this paper, we provide a 3D spherical-cap fitting procedure (3D-SCFP) with many advantages and capabilities. The procedure integrates powerful feature extraction method, namely the Circle Hough Transform. Through this method, the data points of the truncated nanodroplets and the overlapped microcaps can be accurately separated. Then the procedure applies 3D fits to the nanodroplet and the microstructure separately. The details of the procedure and the MATLAB program are provided. We also provide a comparison between 3D-SCFP and the often-used 2D cross-sectional profile fitting method by applying them on two AFM images: one with an isolated nanodroplet and the other complex one with two truncated nanodroplets sitting on a nanocap. The  2D fits
have the following shortcomings: (i) the uncertainty caused by the arbitrary selection of the cross-sectional profile; (ii) the limited number  of data points taken for the fitting and 
the too large statistical weight of data points close to the top of the spherical cap; and (iii) inconvenience in the truncated nanodroplet case lies in the small amount of fitted data points, the artificial error of the cross section selection and the possible distortion of the extracted profile nearby the contact line.
The first two shortcomings lead the inconsistent results of 2D cross-sectional profile fitting method. 
When a required cross section is not along the pixel array in the truncated nanodroplet case, the difficulty of extracting an accurate cross-sectional profile makes the last shortcoming more obvious. 
However, all the defects above are overcome in the 3D-SCFP as it considers all the valid data points and fits
 them in 3D.
The comparison  confirms that for both cases.
Therefore the 3D-SCFP is strongly preferable. 

We expect our  3D spherical-cap fitting procedure to find many applications in image analysis. 
It may be particularly useful for 
 case 2 in which  (nano)droplets overlap with a (nano)cap (or other droplets), as then the potential of the procedure
 is fully explored. This case  2 is omnipresent in nature \cite{extrand2012,zhang201688,neeson2012,mahadevan2002} and (nano)technology \cite{lohse2015rmp,peng2015spontaneous,peng2016}. 
The morphology of this case with unconventional shapes can be used as effective templates \cite{guzowski2015,kraft2009,kraft2011} in fabrication. 
In  recently published work 
 \cite{park2016} it was
  reported that the droplet nucleation on convex shape, such as microcaps, significantly enhances the  heat transfer. 
That paper underlines various 
 important applications of convex shapes in water collection and heat transfer connected with phase transitions. 
 One famous biomimetic structure is the back of a desert beetle, where water droplets form on the spherical lumps of the hydrophilic domain.
Another famous biomimetic concept is the `Lotus Effect' due to  micro- and nanoscopic architecture under a droplet \cite{latthe2014,ma2009,ensikat2011}.
A mimicked lotus-leaf surface is fabricated by making  structures with convex shapes on top of a  surface \cite{liu2006}.
The morphology of droplets on caps with unconventional shapes also provides effective templates \cite{guzowski2015,kraft2009,kraft2011} in fabrication of sophisticated microparticles. 
How to precisely extract morphologic characteristics of truncated spherical caps from experimental data is crucial for all these cases.
The method and codes created in this paper are readily implemented for imaging analysis of similar (nano)structures, and are accessible to all users through the supplementary materials online.

\begin{addendum}
 \item We thank Ivan Devi\'{c} for discussion and testing of 3D spherical-cap fitting procedure.
We gratefully acknowledge NWO because of support through the MCEC Zwaartekracht program.
D.L. in addition acknowledges the support from an ERC Advanced Grant and X.H.Z the support from Australian Research Council (FT120100473, DP140100805).
H. T. thanks for the financial support from the China Scholarship Council (CSC, file No. 201406890017).
 \item[Competing Interests] The authors declare that they have no
competing financial interests.
 \item[Correspondence Email:] h.tan@utwente.nl or xuehua.zhang@rmit.edu.au or d.lohse@utwente.nl.
\end{addendum}

\section*{Appendix A. Data points division}
The projector of the contour of the truncated nanodroplet on the top surface of the microcap (the black solid line) is a part of an ellipse. It is within the overlapping region of two detected circles. In the ``Objective detection" step of 3D-SCFP, we prepare data points inside the microcap-circle and outside the nanodroplet-circle (Data group 1) for 3D fit of the microcap, while the data points inside the nanodroplet-circle and outside the microcap-circle (Data group 2-1) for 3D fit of the nanodroplet. 

\begin{figure*}
\centering
\resizebox{0.5\linewidth}{!}{%
  \includegraphics{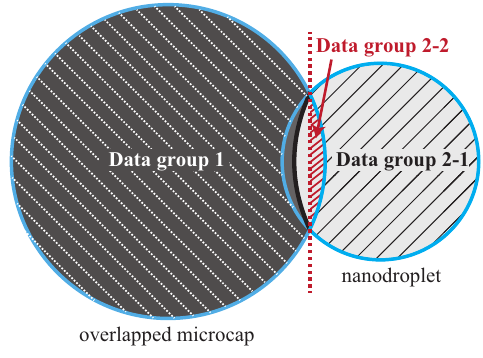}
}
\caption{Schematic diagram of dividing the joint data points of a nanodroplet (the light grey geometry) and the overlapped microcap (the dark grey geometry). Two blue circles, detected by Circle Hough Transform, provide two regions: Data group 1 (with the pattern of white diagonal dot-line) and Data group 2-1 (with black-diagonal-line pattern). The red dash-line passes the intersection points and defines the region of Data group 2-2 (with red-line pattern). 
}
\label{fig:appA}       
\end{figure*}

In order to also consider contributions from the data points in the overlapping region, we divide the region into two parts with a straight line (the red dash-line). The line passes through two intersection points of two circles. Once the nanodroplet is recognised in ``Objective identification" step, we can determine the part that all belongs to the nanodroplet. The part (Data group 2-2) is labeled with a red-line pattern as displayed in figure~\ref{fig:appA}. Hence, both the data group 2-1 and the data group 2-2 are used in 3D fit of the overlapping nanodroplet in the final step of 3D-SCFP.

\section*{Appendix B. Cross-sectional profile extraction}
\begin{figure*}
\centering
\resizebox{0.5\linewidth}{!}{%
  \includegraphics{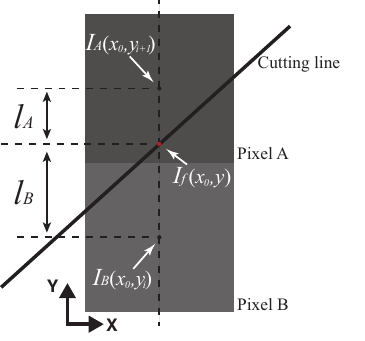}
}
\caption{Schematic diagram of extracting an arbitrary line from a pixel-array/data-matrix. Two big solid squares represent two pixel points A and B with corresponding recorded signal (sample height in the pixel centre) $I_A$ and $I_B$. When a line doesn't pass through the centre of the pixel, the grey values is $I_f$.}
\label{fig:appB}       
\end{figure*}

An AFM image is a raster image.
Each pixel has a relative x-y position in the image. 
The pixel intensity correspond to the sample height (or cantilever deflection) at that location.
When we extract a cross-sectional profile along any direction except $0^o$, $45^o$ or $90^o$, it is common to come across the situation shown in figure.~\ref{fig:appB}.
The cutting line of the cross section does not pass through any pixel centres in a pixel column (row).
In order to obtain the height value $I_{f}$ at position $(x_0,y)$, we assume there existing a linear variation between the height values at positions $(x_0,y_i)$ and $(x_0,y_{i+1})$.
Hence, $I_{f}$ is calculated as,
\begin{equation}
I_f=\frac{l_BI_A+l_AI_B}{l_A+l_B},
\end{equation}
where $I_A$ and $I_B$ are the distances from the investigating point to the neighbour pixel centres. 
The AFM image used in figures~\ref{fig:res2} has a high resolution, so the linear fit is precise enough.
For a low resolution AFM image, it is required to apply a high order fit which considers contributions from more neighbour pixels.

\section*{Appendix C. Matlab Codes}
Matlab codes of 3D spherical-cap fitting procedure are available for any applications and future development for any purpose. 
A resulting publications should cite this paper.


%
%
%


\end{document}